\documentclass[aps,prl,twocolumn,nofootinbib,showpacs,psfig,superscriptaddress,longbibliography]{revtex4-2}

\usepackage{color}
\usepackage[colorlinks=true,linkcolor=blue,citecolor=blue]{hyperref}
\usepackage{hyperref}
\usepackage{amsmath}
\usepackage{amssymb}
\usepackage{graphicx}
\usepackage{epsfig}
\usepackage{dcolumn}
\usepackage{bm}
\usepackage{bm}
\usepackage{blindtext}
\usepackage{natbib}
\usepackage{booktabs}
\usepackage{mathrsfs}
\usepackage{epstopdf}

\usepackage{textcomp}
\usepackage{times}
\usepackage{float}
\usepackage{latexsym,amsmath,amssymb,bm,euscript}
\usepackage{subfigure}
\usepackage{ulem}

\usepackage{graphicx}
\usepackage{ulem}
\def\be{\begin{equation}}
\def\ee{\end{equation}}
\def\bea{\begin{eqnarray}}
\def\eea{\end{eqnarray}}

\DeclareGraphicsExtensions{.pdf, .png, .jpg, .eps}

\begin{document}

\title{NMR evidence of spin supersolid and Pomeranchuk effect behaviors in the triangular-lattice antiferromagnet Rb$_2$Ni$_2$(SeO$_3$)$_3$}

\author{Ying Chen}
\thanks{These authors contributed equally to this study.}
\affiliation{Department of Physics and Beijing Key Laboratory of Opto-electronic Functional Materials $\&$ Micro-nano Devices, Renmin University of China, Beijing, 100872, China}

\author{Zhanlong Wu}
\thanks{These authors contributed equally to this study.}
\affiliation{Department of Physics and Beijing Key Laboratory of  Opto-electronic Functional Materials $\&$ Micro-nano Devices, Renmin University of China, Beijing, 100872, China}

\author{Xuejuan Gui}
\thanks{These authors contributed equally to this study.}
\affiliation{Department of Physics and Beijing Key Laboratory of  Opto-electronic Functional Materials $\&$ Micro-nano Devices, Renmin University of China, Beijing, 100872, China}

\author{Guijing Duan}
\thanks{These authors contributed equally to this study.}
\affiliation{Department of Physics and Beijing Key Laboratory of Opto-electronic Functional Materials $\&$ Micro-nano Devices, Renmin University of China, Beijing, 100872, China}

\author{Shuo Li}
\affiliation{Institute of Physics, Chinese Academy of Sciences, and Beijing National Laboratory for Condensed Matter Physics, Beijing 100190, China}  

\author{Xiaoyu Xu}
\affiliation{Department of Physics and Beijing Key Laboratory of  Opto-electronic Functional Materials $\&$ Micro-nano Devices, Renmin University of China, Beijing, 100872, China}

\author{Kefan Du}
\affiliation{Department of Physics and Beijing Key Laboratory of  Opto-electronic Functional Materials $\&$ Micro-nano Devices, Renmin University of China, Beijing, 100872, China}

\author{Xinyu Shi}
\affiliation{Department of Physics and Beijing Key Laboratory of  Opto-electronic Functional Materials $\&$ Micro-nano Devices, Renmin University of China, Beijing, 100872, China}

\author{Rui Bian}
\affiliation{Department of Physics and Beijing Key Laboratory of  Opto-electronic Functional Materials $\&$ Micro-nano Devices, Renmin University of China, Beijing, 100872, China}

\author{Xiaohui Bo}
\affiliation{Department of Physics and Beijing Key Laboratory of  Opto-electronic Functional Materials $\&$ Micro-nano Devices, Renmin University of China, Beijing, 100872, China}

\author{Guochen Liu}
\affiliation{Department of Physics and Beijing Key Laboratory of  Opto-electronic Functional Materials $\&$ Micro-nano Devices, Renmin University of China, Beijing, 100872, China}

\author{Jun Luo}
\affiliation{Institute of Physics, Chinese Academy of Sciences, and Beijing National Laboratory for Condensed Matter Physics, Beijing 100190, China}  

\author{Jie Yang}
\affiliation{Institute of Physics, Chinese Academy of Sciences, and Beijing National Laboratory for Condensed Matter Physics, Beijing 100190, China}  

\author{Yi Cui}
\email{cuiyi@ruc.edu.cn}
\affiliation{Department of Physics and Beijing Key Laboratory of  Opto-electronic Functional Materials $\&$ Micro-nano Devices, Renmin University of China, Beijing, 100872, China}
\affiliation{Key Laboratory of Quantum State Construction and Manipulation (Ministry of Education),
Renmin University of China, Beijing, 100872, China}

\author{Rui Zhou}
\email{rzhou@iphy.ac.cn}
\affiliation{Institute of Physics, Chinese Academy of Sciences, and Beijing National Laboratory for Condensed Matter Physics, Beijing 100190, China}

\author{Jinchen Wang}
\email{jcwang\_phys@ruc.edu.cn}
\affiliation{Department of Physics and Beijing Key Laboratory of  Opto-electronic Functional Materials $\&$ Micro-nano Devices, Renmin University of China, Beijing, 100872, China}
\affiliation{Key Laboratory of Quantum State Construction and Manipulation (Ministry of Education),
Renmin University of China, Beijing, 100872, China}

\author{Rong Yu}
\email{rong.yu@ruc.edu.cn}
\affiliation{Department of Physics and Beijing Key Laboratory of  Opto-electronic Functional Materials $\&$ Micro-nano Devices, Renmin University of China, Beijing, 100872, China}
\affiliation{Key Laboratory of Quantum State Construction and Manipulation (Ministry of Education),
Renmin University of China, Beijing, 100872, China}

\author{Weiqiang Yu}
\email{wqyu\_phy@ruc.edu.cn}
\affiliation{Department of Physics and Beijing Key Laboratory of  Opto-electronic Functional Materials $\&$ Micro-nano Devices, Renmin University of China, Beijing, 100872, China}
\affiliation{Key Laboratory of Quantum State Construction and Manipulation (Ministry of Education),
Renmin University of China, Beijing, 100872, China}

\begin{abstract}
We performed $^{85}$Rb nuclear magnetic resonance (NMR) measurements on the $S$ = 1 bilayer triangular-lattice antiferromagnet Rb$_2$Ni$_2$(SeO$_3$)$_3$ in magnetic fields up to 26~T. In the field range from 3~T to 26~T, the NMR spectral lines split and their respective spectral weight ratios reveal the existence of the magnetic up-up-down (UUD) phase, although the 1/3-plateau phase is only reached at fields above 16~T. Two distinct gapless regimes are further identified: one at low fields and low temperatures, and the other at high fields and high temperatures, consistent with the spin supersolid Y and V phases. Notably, the UUD-V phase boundary exhibits a negative slope in $dT/dH$, where the supersolid phase is located at temperatures above the solid phase due to strong low-energy spin fluctuations.
\end{abstract}

\maketitle


{\bf Introduction.--}
Frustrated systems exhibit a variety of novel physical properties driven by their  macroscopic degeneracy~\cite{2024_PRM_Aczel,2019_PNAS_Yoshimitsu,2015_PPP_Starykh,1985_PRB_Yosefin,1978_JPC_Derrida}. 
The triangular lattice antiferromagnets (TLAFMs)  represent such a celebrated example and have been extensively studied in both quantum and classical regimes~\cite{1985_JPSJ_Kawamura,2004_JPSJ_Yoshikawa,2011_PRB_Seabra,2013_IOP_Hiroyuki,2014_PRL_Yamamoto,2016_JMMM_Farnell}.
They provide a versatile platform for exploring novel phenomena, including magnetic plateau~\cite{1991_JPCM_Chubukov,2009_JPCM_Farnell,2013_PRL_Hiroyuki}, quantum spin liquid~\cite{2010_nature_Balents,2016_NP_Bella,2018_npj_Adam,2019_NP_Bruce}, and spin supersolid (SS)~\cite{2009_PRL_Wang,2009_PRB_Jiang,2010_PRL_Heidarian}. 
Furthermore, these phases are highly sensitive to and can be tuned by various perturbations, such as spin anisotropy, longer-range interactions, 
magnetic fields, impurities, and lattice deformations, under the interplay of thermal and quantum fluctuations
~\cite{2021_PT_Akai,2017_PRB_Hiroyuki,2014_PRL_Yamamoto,2009_PRL_Chris,2004_JPSJ_Yoshikawa,1995_PU_Dotsenko,1994_HI_Bruce}. As a unique quantum state, the SS extends the original concept of supersolid~\cite{1970_PRL_Leggett} to magnetic systems. It arises  from spontaneous breaking of both 
spin rotational and lattice symmetries, and is characterized as a planar order coupled to an UUD spin pattern in TLAFMs.  Experimental evidence of spin supersolidity has only been reported in several $S=1/2$ triangular-lattice compounds, such as Na$_2$BaCo(PO$_4$)$_2$~\cite{2024_nature_sunpeijie, 2025_PRB_Xu, 2022_PNAS_JMSheng}, K$_2$Co(SeO$_3$)$_2$~\cite{2024_PRL_zhu,2024_arxiv_Chen}, and Rb$_2$Co(SeO$_3$)$_2$~\cite{2025_arxiv_cui}. 

Recently, a class of bilayer triangular lattice materials has attracted significant attention~\cite{1998_PRL_Singh,2018_PRB_Jozef,2024_PRL_chengang,2020_PRB_zhongruidan,2024_CM_XuXianghan,2024_PRB_MeiJW,2024_PRB_WangJ.F}. In these systems, the interplay of intralayer geometric frustration and  
interlayer couplings
 enhances quantum fluctuations, resulting in emergent critical behaviors under a magnetic field, such as the Kawamura universality class~\cite{1998_PRL_Singh,1992_jpsj_Kawamura}, Potts criticality~\cite{2018_PRB_Jozef,1982_RMP_FYWu}, and Berezinskii-Kosterlitz-Thouless (BKT) phase transitions~\cite{2024_PRL_chengang}, as well as candidate quantum spin liquids~\cite{2020_PRB_zhongruidan}, which extend beyond the traditional frameworks of antiferromagnets. An open and intriguing question is whether exotic phases can emerge in $S=1$ TLAFMs, and the Ising anisotropic bilayer TLAFM K$_2$Ni$_2$(SeO$_3$)$_3$ with weak bilayer interactions ~\cite{2024_PRB_MeiJW} offers a case study to address this. It not only exhibits a 1/3-like magnetization plateau between 18 T and 26 T~\cite{2024_PRB_WangJ.F} but also shows
double magnetic transition lines in both magnetic susceptibility and heat capacity measurements~\cite{2024_PRB_MeiJW,2024_PRB_WangJ.F}. These features resemble those found in easy-axis anisotropic TLAFMs and hint at the stabilization of UUD and SS phases~\cite{1985_JPSJ_Kawamura}.

\begin{figure}[t]
\includegraphics[width=8.5cm]{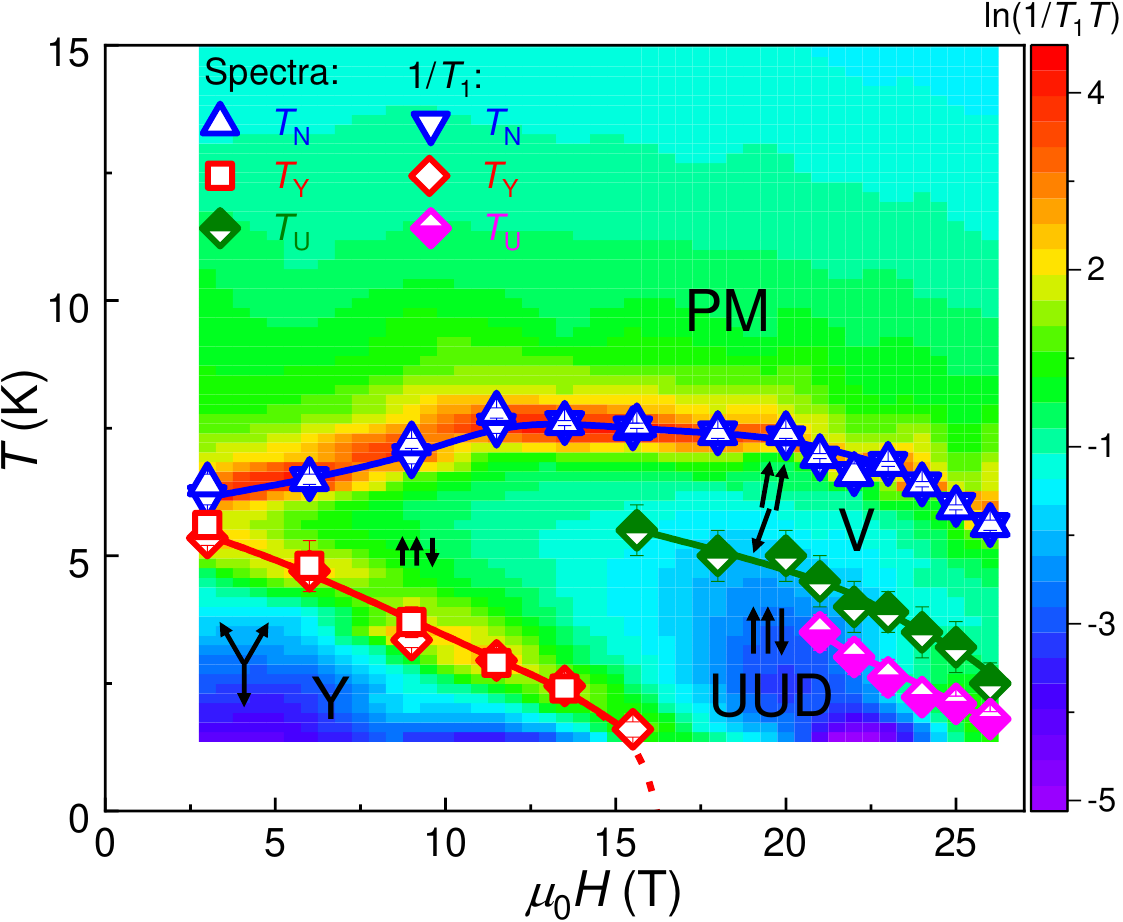}
\caption{\label{pd}{\bf Phase diagram of Rb$_2$Ni$_2$(SeO$_3$)$_3$.}
Heat map represents the contour plot of $1/T_1T$ data. $T_{\rm N}$, $T_{\rm Y}$, and $T_{\rm U}$  represent 
the phase boundaries among the PM, UUD, Y, and V phases, which are determined from the spectral and the spin-lattice relaxation rate data as specified. Solid and dotted lines are guides to the eyes. 
}
\end{figure}

In this letter, we report NMR measurements on Rb$_2$Ni$_2$(SeO$_3$)$_3$, an isostructural compound to K$_2$Ni$_2$(SeO$_3$)$_3$, in fields up to 26~T. The main results are summarized in the phase diagram of Fig.~\ref{pd}, which is overlaid on the spin-lattice relaxation rate data $1/T_1$. A dome-shaped N\'{e}el transition  $T_{\rm N}$ is determined from both the NMR line splitting and the peaks in $1/T_1$. Below $T_{\rm N}$, we resolve two additional transition lines at $T_{\rm Y}$ and $T_{\rm U}$, respectively.  Our spectral analysis in the intermediate region intercepting between the $T_{\rm Y}$ and $T_{\rm U}$ lines reveals the UUD phase, though the 1/3-magnetization plateau phase is only achieved at field above 16~T~\cite{2024_PRB_WangJ.F}. Gapless behavior is observed below the $T_{\rm Y}$ line and above the $T_{\rm U}$ line, providing strong evidence for two SS phases. Remarkably, the V-UUD phase boundary exhibits a negative slope $dT/dH$, signifying a counterintuitive transition from a low-symmetry to a high-symmetry phase upon cooling. This observation is consistent with a magnetic analogue of the Pomeranchuk effect~\cite{1950_ZETF_pomeranchuk,1997_RMP_Lee,1997_RMP_Richardson,2025_PRB_Xtao}. Our study elucidates novel emergent quantum phenomena in the $S=1$ bilayer antiferromagnet.

\begin{figure}[t]
\includegraphics[width=8.5cm]{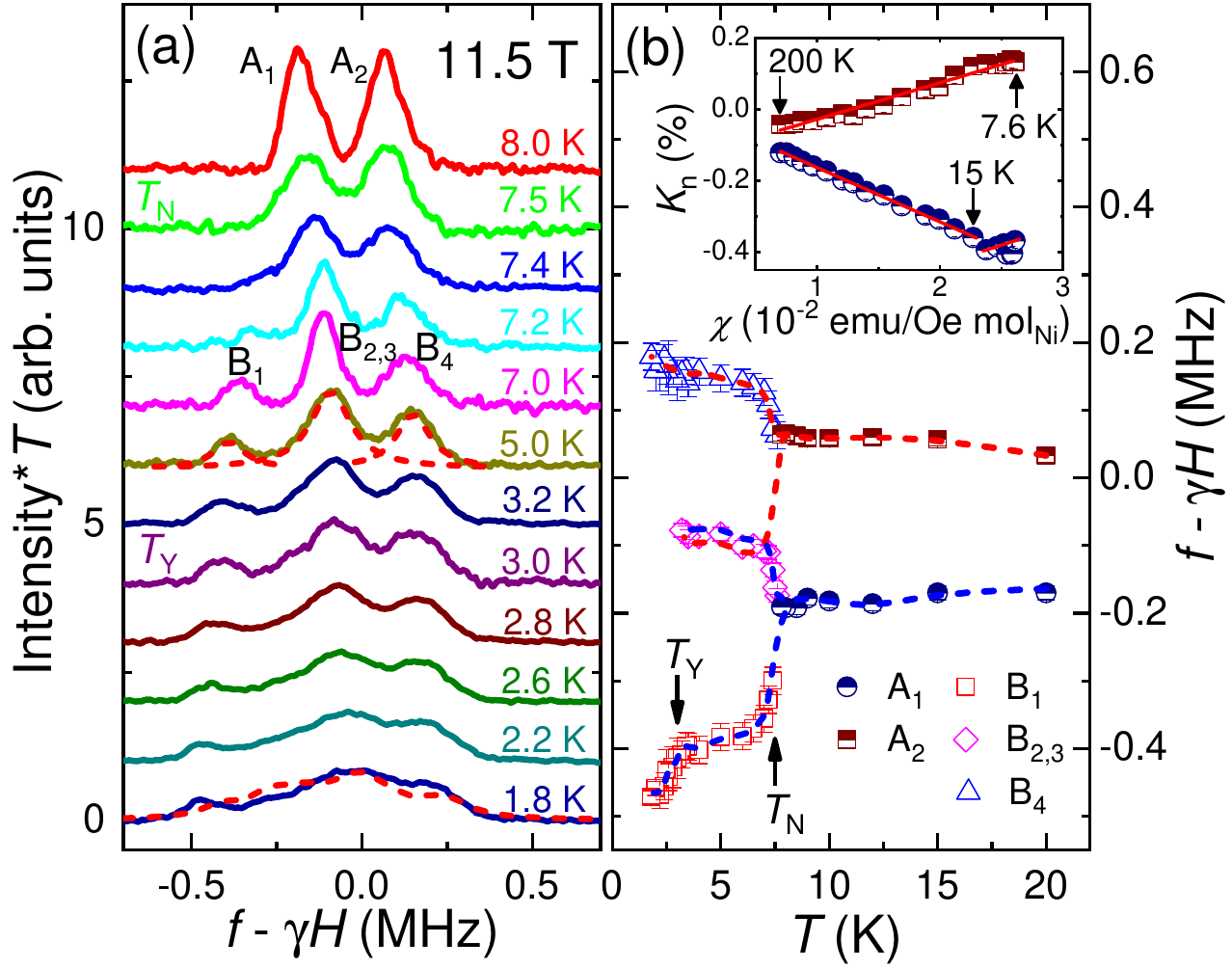}
\caption{\label{specl}{\bf Low-field NMR spectra.}
(a) $^{85}$Rb center lines measured at 11.5~T with decreasing temperatures. Vertical data offsets are applied for clarity. A$_1$ and A$_2$ are two peaks observed at and above 7.5~K, and B$_1$, B$_{2,3}$, and B$_4$ represent peaks resolved at lower temperatures. A three-Lorentzian fit is applied to the data at 5~K, with the relative spectral weight fixed at a ratio of 1:3:2. The simulated spectrum of the Y phase is plotted at 1.8~K (see text).  
(b) Frequencies of all resolved peaks as functions of temperatures. $T_{\rm {N}}$ and $T_{\rm Y}$ denote transition temperatures determined from the change of the frequency in the peaked spectra. Dashed lines are guides to the eyes.
Inset: High-temperature $K_{\rm n}$ of A$_1$ and A$_2$ plotted against magnetic susceptibility. The solid straight lines are linear function fits to the data to obtain the hyperfine coupling constant (see text). 
}
 \end{figure}

{\bf Low-field NMR spectra and double  phase transitions.--}  
Detailed material and measurement techniques are described in the Supplementary Materials (SM)~\cite{SM}. The central NMR lines of $^{85}$Rb are shown in Fig.~\ref{specl}(a) at different temperatures under a low field of 11.5~T. 

A$_1$ and A$_2$, with a spectral weight ratio of about 1:1, are resolved. They are attributed to two inequivalent $^{85}$Rb sites caused by randomly half-occupied Se(II) ions  [see Sec.~S1 in the SM~\cite{SM}]. 

Below 7.5~K, both lines split: A$_1$ splits into two peaks, labeled B$_1$ and B$_3$, and A$_2$ splits into B$_2$ and B$_4$. Since
B$_2$ and B$_3$ overlap, only three peaks are observed. This line splitting indicates a phase transition from the paramagnetic (PM) state to a symmetry-breaking state upon cooling. Below 3~K, the spectra exhibit significant broadening, making it difficult to resolve peaks B$_2$ and B$_3$, while B$_1$ and B$_4$ remain resolvable. The frequencies of the respective peaks were then collected and plotted as functions of temperature in Fig.~\ref{specl}(b). The calculated Knight shifts $K_{\rm n}(T)$ for A$_1$ and A$_2$ at high temperatures are plotted against the magnetic susceptibility $\chi(T)$ in the inset of Fig.~\ref{specl}(b), with temperature as an implicit variable. From linear fits, the hyperfine coupling constants are obtained as $A_{\rm hf} = 0.057(2)$ T/$\mu_{\rm B}$ for A$_2$, and $A_{\rm hf} = -0.085(1)$ T/$\mu_{\rm B}$ and $A_{\rm hf} = 0.062(2)$ T/$\mu_{\rm B}$ for A$_1$ at temperatures above and below 15~K, respectively.

Below 7.5~K, the frequencies of B$_{2,3}$ and B$_4$ exhibit a sharp increase and then level off, while that of B$_1$ drops rapidly before leveling off. Below 3~K, the frequencies of B$_1$ (B$_4$) exhibit a kinked decrease (increase) upon further cooling. These changes in the spectra clearly signal two successive phase transitions upon cooling, at $T_{\rm N}=7.5$ K and $T_{\rm Y}=3$ K, corresponding to the PM-UUD and UUD-Y phase transitions, respectively, as will be substantiated below.

The spectrum at 5~K is fitted using a three-Lorentzian function, as shown in Fig.~\ref{specl}(a). The fit yields a relative spectral weight ratio of about 1:3:2 from left to right. This distribution is consistent with the UUD phase given the two inequivalent $^{85}$Rb sites: The UUD spin configuration, with two up spins and one down spin, predicts a 1:2 spectral weight ratio for each $^{85}$Rb site. The observed 1:3:2 ratio arises from an accidental overlap of one peak from each $^{85}$Rb site (B$_2$ and B$_3$), as supported by the simulated spectra in Sec.~S3 of the SM~\cite{SM}.

We also performed spectral analysis on the data at 1.8~K [Fig. S3 in the SM~\cite{SM}], and found that good agreement between the simulated and measured data could be achieved only when we assumed the system to be in the Y phase [Fig.~\ref{specl}(a)], in which both in-
and out-of-plane order contribute to the line splitting via the dipolar-type hyperfine  
coupling to the ligand sites. The spectrum shows significant broadening in each line compared to that above 3~K, 
indicating a change in magnetic structure from the UUD to the supersolid Y phase with 
the development of the in-plane AFM order. The gapless excitations, further support for the supersolid nature, are confirmed by the 
power-law temperature dependence of the $1/T_1$ data discussed below. The transition temperatures $T_{\rm {Y}}$ and $T_{\rm N}$ 
determined from the spectra for fields up to 15.6~T  are marked in the phase diagram of Fig.~\ref{pd}.
 
\begin{figure}[t]
\includegraphics[width=8.5cm]{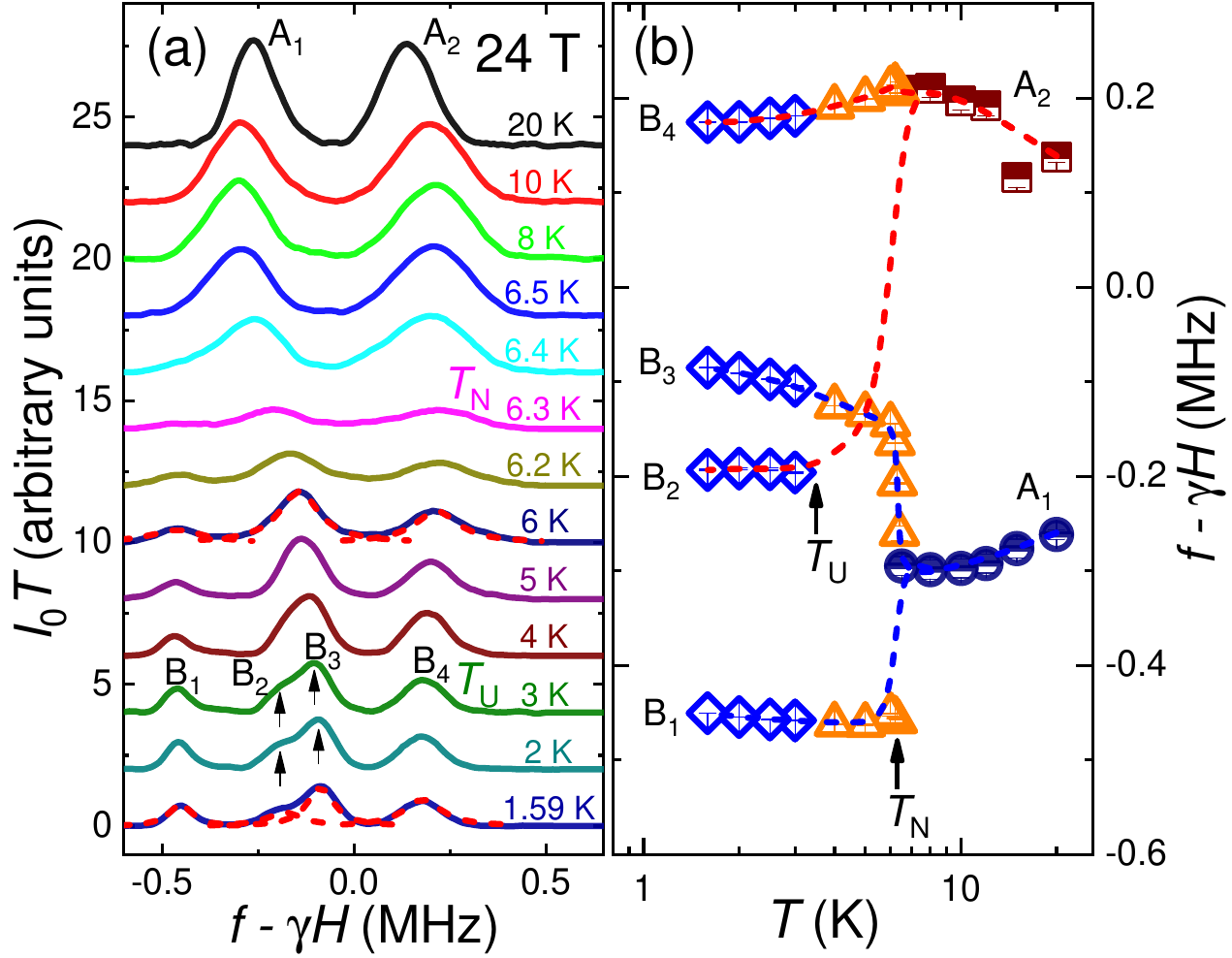}
\caption{\label{spech} {\bf High-field NMR spectra.}
(a) Spectra at 24~T measured with decreasing temperatures. Data are shifted vertically for clarity. B$_1$ to B$_4$ label the peaks resolved at low temperatures. Dashed lines are three-Lorentzian function fit applied to data at 6~K with fixed spectral weight ratio of 1:3:2 from left to the right, and four-Lorentzian function fit to data at 1.59~K with fixed ratio of 1:1:2:2.
(b) Resonance frequencies of the respective spectral peaks measured at 24~T as functions of temperatures. Dash lines are guides to the eyes.
}
 \end{figure}

{\bf High-field spectra and successive phase transitions.--}
Figure~\ref{spech}(a) displays the spectra measured at a high field of 24~T. Two NMR lines due to inequivalent $^{85}$Rb sites are still seen in the PM phase above 6.3~K. Upon cooling, a three-peak feature develops, indicating the onset of magnetic ordering. Below 4~K, a four-peak feature is clearly resolved. The corresponding frequencies of these peaks are then plotted as functions of temperature 
in Fig.~\ref{spech}(b). 
Two magnetic transitions corresponding to significant frequency changes are then identified at $T_{\rm N}=$ 6.3~K and $T_{\rm U}=$ 3.5~K, 
respectively. 

The determined $T_{\rm {N}}$ and $T_{\rm U}$ values are summarized in the phase diagram of Fig.~\ref{pd}.  Here $T_{\rm U}$ can be understood as the  transition temperature from the supersolid V phase to the UUD phase upon cooling. To see this, we examine the low-temperature spectra in Fig.~\ref{spech}(a). The spectrum at 6~K follows a three-Lorentzian function fit with a spectral ratio of 1:3:2 for peaks from left to right, suggesting the breaking of $c$ axis lattice symmetry in a way similar to the UUD phase. Upon cooling below about 3~K, the spectrum splits into four peaks, with A$_1$ splitting into B$_1$ and B$_3$, and A$_2$ into B$_2$ and B$_4$, respectively. The split spectrum can be fitted by a four-Lorentzian function with a spectral weight ratio close to 1:1:2:2, as shown in the 1.59~K data in Fig.~\ref{spech}(a).  This is consistent with the UUD phase, with B$_3$ (B$_1$) and B$_4$ (B$_2$) representing contributions from up (down) spins with a population ratio of 2:1. 

Our spectral simulations further confirm that the three-peak spectra between 3~K and 6~K are in good agreement with the V phase [Fig.~S3(g) in the SM~\cite{SM}], while the four-peak feature is consistent with the UUD phase with an enhanced staggered magnetic moment compared to the low-field data [Fig.~S4 in the SM~\cite{SM}]. In fact, the $1/T_1$ data shown below clearly resolve gapless excitations at temperatures between $T_{\rm N}$ and $T_{\rm U}$, and a gapped one at temperatures below $T_{\rm U}$. Therefore, $T_{\rm N}$ and $T_{\rm U}$
are understood as successive phase transitions from the PM to the supersolid V phase, then to the magnetic UUD phase.

\begin{figure}[t]
\includegraphics[width=7.5cm]{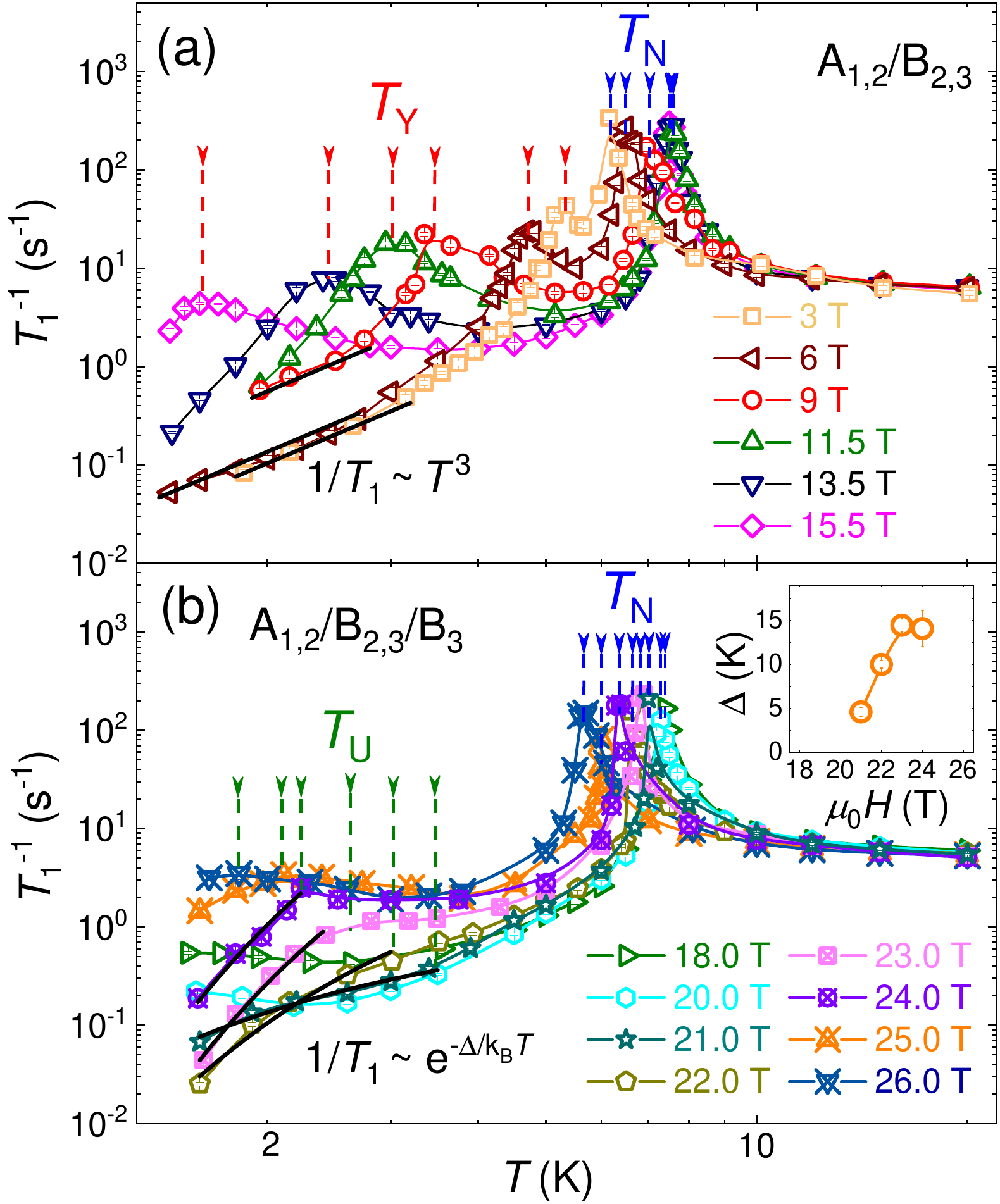}
\caption{\label{slrr} {\bf Spin-lattice relaxation rates.}
(a) Low-field $1/T_1$ data as functions of temperatures. The blue and red arrows mark the peak positions labeled by $T_{\rm N}$ and $T_{\rm Y}$, respectively. Solid straight lines are guides to the power-law scaling 1/$T_1{\sim}T^3$. (b) High-field $1/T_1$ as functions of temperatures. Blue and green arrows mark the peak and the hump positions, labeled by $T_{\rm N}$ and $T_{\rm U}$ respectively. The solid lines are empirical function fits to $1/T_1{\sim}e^{-{\Delta}/{k_{\rm{B}}T}}$ with an energy gap $\Delta$. Inset: The determined $\Delta$ as a function of field. 
}
 \end{figure}

{\bf Spin-lattice relaxation rate.--}
The low-energy spin fluctuations were probed by $1/T_1$. Figure~\ref{slrr}(a) and (b) show the temperature dependence of $1/T_1$ at low  and high fields, respectively,  measured on different spectral peaks as defined in Fig.~\ref{specl} and Fig.~\ref{spech}.
Note that $1/T_1$ on A$_1$ is identical to that on A$_2$. 

At low fields, as the temperature decreases from 20~K to 10~K, $1/T_1$ increases slowly, evidencing the onset of low-energy spin fluctuations; whereas below 10~K, $1/T_1$ increases dramatically, forming a sharp peak at $T_{\rm N}$. This demonstrates the onset of spin fluctuations as the system approaches magnetic ordering. Upon further cooling, a broad peak is observed at a lower temperature $T_{\rm Y}$, indicating the emergence of a second magnetic transition. $T_{\rm {N}}$ and $T_{\rm Y}$ determined from the $1/T_1$ data are then added 
to the phase diagram [Fig.~\ref{pd}] and they are consistent with those determined from the NMR spectra.
Below $T_{\rm Y}$, $1/T_1$ drops gradually with decreasing temperature and follows a power-law behavior $1/T_1{\sim}T^3$ for fields between 3 and 9~T. This power-law temperature dependence is indicative of gapless excitations associated with the broken in-plane continuous symmetry of the supersolid Y phase~\cite{2025_PRB_Xu}.

At high fields, $1/T_1$ also increases upon cooling and develops a sharp peak at $T_{\rm N}$. Below $T_{\rm N}$, $1/T_1$ drops quickly and levels off over a broad temperature range. Then it forms a hump feature at a lower temperature marked $T_{\rm U}$. $T_{\rm N}$ and $T_{\rm U}$ determined from $1/T_1$ are then added to the phase diagram in Fig.~\ref{pd}. $T_{\rm N}$ determined from $1/T_1$ are in good 
agreement with that determined from the spectra.  However, $T_{\rm U}$ determined from $1/T_1$ are slightly lower than that 
from the spectra.

For $T<T_{\rm U}$, $1/T_1$ drops rapidly and follows a thermal activation behavior $1/T_1=ae^{-{\Delta}/{k_{\rm{B}}T}}$. 
The gap $\Delta$ extracted from the function fit at different fields is summarized in the inset of Fig.~\ref{slrr}(b). $\Delta$ increases with field and reaches a maximum of 15~K at 23~T. Such a large gap value provides strong evidence for the stabilization of a UUD phase in the high-field, low-temperature regime. By comparison, $1/T_1$ exhibits a gradual decrease
over a large temperature regime between $T_{\rm N}$ and $T_{\rm U}$ across all measured fields, highly suggestive of strong low-energy spin fluctuations in this intermediate temperature regime, which imply the emergence of low-lying excitations contributing to magnetic entropy.

Note that our analysis has already shown that the NMR spectra in this regime are consistent with the supersolid V phase. Taken together, and given the Ising anisotropy of the system~\cite{2024_PRB_MeiJW}, the gapless excitations provide compelling evidence for the supersolid V phase originally proposed for TLAFMs~\cite{2014_PRL_Yamamoto}.

{\bf Phase diagram and discussions.--} 
From the data presented above, we construct a complete magnetic phase diagram consisting of PM, UUD, Y, and V phases.
To precisely understand the nature of the phases, the phase diagram is overlaid on top of the colored contour map of $1/T_1T$ in Fig.~\ref{pd}. It reveals a single dome in $T_{\rm N}$ as a function of field which is consistently determined from both the spectral splitting and the sharp peak in $1/T_1$. The $T_{\rm Y}$ and $T_{\rm U}$ lines are separate and nearly parallel to each other, enclosing the UUD phase within an intermediate ($H$, $T$) regime. 
Here we note that besides the lattice symmetry breaking and gapless excitations identified by current NMR measurement, signatures from other probes, such as  
the pseudo-Goldstone mode---a gap arising from the out-of-plane order but reduced by the interplay with the in-plane order~\cite{2024_PRL_zhu}, are desirable to fully characterize the SS phase in the current compound. 

While our $T_{\rm N}$ line agrees with that reported in the isostructural compound K$_2$Ni$_2$(SeO$_3$)$_3$ (up to 9 T)~\cite{2024_PRB_MeiJW}, Rb$_2$Ni$_2$(SeO$_3$)$_3$ shows a much stronger suppression of T$_{\rm Y}$ with magnetic field. 

The staggered magnetization from 2.5~K to 6~K, calculated from the NMR line splitting, increases with field up to 16~T and then levels off (see Fig.~S4 in the SM~\cite{SM}). Although the magnetic structure follows the UUD pattern at fields below 16~T, the staggered moment remains small. This indicates the existence of strong fluctuations in the low-field UUD phase, as supported by the absence of gapped behavior at temperatures between $T_{\rm N}$ and $T_{\rm Y}$ [see $1/T_1$ data in Fig.~\ref{slrr}(a)]. The magnetic plateau phase,
as characterized by the gapped behavior [Fig.~\ref{slrr}(b)], is only achieved at fields above 18~T, consistent with the reported 
1/3 magnetic plateau~\cite{2024_PRB_WangJ.F}.

As summarized in Fig.~\ref{pd}, the characteristic temperature $T_{\rm U}$ determined from $1/T_1$ increases from 1.8~K at 26~T to approximately 3.5~K at 21~T, and eventually vanishes near 20~T. The absence of a sharp peak across $T_{\rm U}$ suggests a first-order transition. This conclusion is further strengthened by the slight discrepancy in $T_{\rm U}$ determined from spectral and $1/T_1$ measurements, which indicates a distribution of time scales in the spin dynamics that manifest differently across different experimental probes.

Strikingly, our phase diagram reveals that the supersolid V phase is stabilized at higher temperatures than the UUD one. This behavior is counterintuitive and, to our knowledge, unique among TLAFMs, since the naive expectation is that a phase with more broken symmetries (the V phase, breaking both spin U(1) and lattice $C_3$ symmetries) would be stabilized at a lower temperature than one with fewer broken symmetries (the UUD phase, breaking only the $C_3$).  Note that a similar cooling-induced symmetry recovery is the Pomeranchuk effect reported in the $^3$He system under pressure, where the ordered solid phase lies at higher temperature than the liquid due to the higher nuclear spin entropy in the solid phase~\cite{1950_ZETF_pomeranchuk,1997_RMP_Lee,1997_RMP_Richardson}. However, we should note the Pomeranchuk effect refers to an entropy-driven mechanism arising from two distinct degrees of freedom~\cite{1950_ZETF_pomeranchuk,1997_RMP_Lee,1997_RMP_Richardson,2025_PRB_Xtao}, whereas the current observation, although in a pure spin system, represents an extension of the concept arising from the competition between out-of-plane order and in-plane order in frustrated Ising magnets.

To understand how the physics analogous to the Pomeranchuk effect takes place in our magnetic system, we examine the entropy in the UUD and V phases.  Usually, $\Delta S<0$ is expected with an increase in spin polarization induced by the field. However, the UUD phase has a hard gap $\Delta$; its low-temperature entropy increases as $S_U\propto e^{-\Delta/T}$, whereas the Goldstone mode associated with the broken spin U(1) symmetry in the supersolid V phase gives an entropy $S_V\propto T^2$. We then expect $\Delta S=S_V - S_U>0$ at low temperatures; {\it i.e.}, the subtle interplay of quantum and thermal fluctuations results in larger entropy in the more ordered V phase! In experiment, the level-off behavior in $1/T_1$ over a large temperature range above $T_{\rm U}$ indicates strong spin fluctuations originating from high entropy in the V phase. Across the first-order UUD-to-V transition,  the Clausius–Clapeyron relation predicts $dT$/$dH$ = $-\Delta$$M$/$\Delta$$S$. Since $\Delta M$ is always positive when increasing the magnetic field, $\Delta S>0$ immediately leads to $dT$/$dH$ $~\textless~$0, corresponding to a negative slope of the UUD-to-V phase boundary in the $H–T$ diagram as observed experimentally (Fig.~\ref{pd}). The above analysis is further supported by our theoretical modeling presented in the SM~\cite{SM}. 

Although our NMR data and theoretical model strongly support the entropy-driven mechanism for the negative-slope phase boundary, high-field specific heat measurements are needed to conclusively establish the magnetic Pomeranchuk effect in this system.
The large entropy effect associated with the Pomeranchuk effect may lead to a large magnetocaloric effect and therefore offer potential applications in magnetic refrigeration.

{\bf Summary.--}
In summary, we have investigated the field-induced phases and related phase transitions in an $S=1$ bilayer TLAFM Rb$_2$Ni$_2$(SeO$_3$)$_3$ by NMR experiments in longitudinal fields up to 26 T. We established the existence of the UUD phase in an intermediate temperature and field regime, and two supersolid phases (denoted as Y and V) surrounding the UUD phase at low and high fields, respectively. The UUD phase is also found to survive at low fields but with strong quantum fluctuations, before reaching the gapped 1/3-magnetization plateau phase.  The V phase is found to appear at a higher temperature than the UUD phase, exhibiting a negative $dT/dH$ phase boundary. Such a negative slope and the existence of strong low-energy spin fluctuations in the V phase provide compelling evidence for an entropy-driven Pomeranchuk effect. Our results provide important insights into how the interplay of quantum and thermal fluctuations affects the emergent quantum states in frustrated spin-1 antiferromagnets.

{\it Acknowledgements.} This work was supported by the National Key Research and Development Program of China (Grants No. ~2023YFA1406500 and No.~2024YFA1409200), National Natural Science Foundation of China (Grants No.~12374156, No.~12134020, No.~12334008, No.~12374142, and No.~12304170), the Scientific Research Innovation Capability Support Project for Young Faculty (Grant No.~ZYGXQNJSKYCXNLZCXM-M26), National Key Research and Development Projects of China (Grant No. 2022YFA1403402) and the Strategic Priority Research Program of Chinese Academy of Sciences (Grant No. XDB1270100). 
This work was supported by the Synergetic Extreme Condition User Facility (SECUF, https://cstr.cn/31123.02.SECUF).

\normalem


\begin{thebibliography}{99}\footnotesize
\itemsep=-1pt plus.2pt minus.2pt

\bibitem{2024_PRM_Aczel}
Brassington A, Huang Q, Aczel A~A and Zhou H~D \href{https://link.aps.org/doi/10.1103/PhysRevMaterials.8.014005}{2024 {\em Phys. Rev. Mater.\/} {\bf 8}(1) 014005 }

\bibitem{2019_PNAS_Yoshimitsu}
Kohama Y, Ishikawa H, Matsuo A, Kindo K, Shannon N and Hiroi Z \href{https://doi.org/10.1073/pnas.1821969116}{2019 {\em Proc. Natl. Acad. Sci.\/} {\bf 116} 10686--10690}

\bibitem{2015_PPP_Starykh}
Starykh O~A \href{https://dx.doi.org/10.1088/0034-4885/78/5/052502}{2015 {\em Rep. Prog. Phys.\/} {\bf 78} 052502 }

\bibitem{1985_PRB_Yosefin}
Yosefin M and Domany E \href{https://link.aps.org/doi/10.1103/PhysRevB.32.1778}{1985 {\em Phys. Rev. B\/} {\bf 32}(3) 1778--1795 \href{https://link.aps.org/doi/10.1103/PhysRevB.32.1778}}

\bibitem{1978_JPC_Derrida}
Derrida B, Vannimenus J and Pomeau Y \href{https://dx.doi.org/10.1088/0022-3719/11/23/019}{1978 {\em J. Phys. C: Solid State Phys.\/} {\bf 11} 4749 }

\bibitem{1985_JPSJ_Kawamura}
Kawamura H and Miyashita S \href{https://doi.org/10.1143/JPSJ.54.4530}{1985 {\em J. Phys. Soc. Jpn.\/} {\bf 54} 4530--4538} 

\bibitem{2004_JPSJ_Yoshikawa}
Yoshikawa S, Okunishi K, Senda M and Miyashita S \href{https://api.semanticscholar.org/CorpusID:15055958}{2004 {\em J. Phys. Soc. Jpn.\/} {\bf 73} 1798--1804 \href{https://api.semanticscholar.org/CorpusID:15055958}}

\bibitem{2011_PRB_Seabra}
Seabra L, Momoi T, Sindzingre P and Shannon N \href{https://link.aps.org/doi/10.1103/PhysRevB.84.214418}{2011 {\em Phys. Rev. B\/} {\bf 84}(21) 214418} 

\bibitem{2013_IOP_Hiroyuki}
Nakayama G, Hara S, Sato H, Narumi Y and Nojiri H \href{https://doi.org/10.1088/0953-8984/25/11/116003}{2013 {\em J. Phys.: Condens. Matter\/} {\bf 25} 116003 }

\bibitem{2014_PRL_Yamamoto}
Yamamoto D, Marmorini G and Danshita I  \href{https://link.aps.org/doi/10.1103/PhysRevLett.112.127203}{2014 {\em Phys. Rev. Lett.\/} {\bf 112}(12) 127203}

\bibitem{2016_JMMM_Farnell}
Götze O, Richter J, Zinke R and Farnell D \href{https://www.sciencedirect.com/science/article/pii/S0304885315305254}{2016 {\em J. Magn. Magn. Mater.\/} {\bf 397} 333--341 ISSN 0304-8853 }

\bibitem{1991_JPCM_Chubukov}
Chubukov A~V and Golosov D~I \href{https://dx.doi.org/10.1088/0953-8984/3/1/005}{1991 {\em J. Phys.: Condens. Matter\/} {\bf 3} 69} 

\bibitem{2009_JPCM_Farnell}
Farnell D~J~J, Zinke R, Schulenburg J and Richter J \href{https://dx.doi.org/10.1088/0953-8984/21/40/406002}{2009 {\em J. Phys.: Condens. Matter\/} {\bf 21} 406002} 

\bibitem{2013_PRL_Hiroyuki}
Susuki T, Kurita N, Tanaka T, Nojiri H, Matsuo A, Kindo K and Tanaka H \href{https://link.aps.org/doi/10.1103/PhysRevLett.110.267201}{2013 {\em Phys. Rev. Lett.\/} {\bf 110}(26) 267201 }

\bibitem{2010_nature_Balents}
Balents L \href{https://www.nature.com/articles/nature08917}{2010 {\em Nature\/} {\bf 464} 199} 

\bibitem{2016_NP_Bella}
Balz C, Lake B, Reuther J, Luetkens H, Sch{\"o}nemann R, Herrmannsd{\"o}rfer T, Singh Y, Nazmul~Islam A, Wheeler E~M, Rodriguez-Rivera J~A, Guidi T, Simeoni G~G, Baines C and Ryll H \href{https://doi.org/10.1038/nphys3826}{2016 {\em Nat. Phys.\/} {\bf 12} 942--949}

\bibitem{2018_npj_Adam}
Banerjee A, Lampen-Kelley P, Knolle J, Balz C, Aczel A~A, Winn B, Liu Y, Pajerowski D, Yan J, Bridges C~A {\em et~al.\/} \href{https://doi.org/10.1038/s41535-018-0079-2}{2018 {\em npj Quant. Mater.\/} {\bf 3} 8 }

\bibitem{2019_NP_Bruce}
Clark L, Sala G, Maharaj D~D, Stone M~B, Knight K~S, Telling M~T, Wang X, Xu X, Kim J, Li Y, Cheong S~W and Gaulin B~D \href{https://doi.org/10.1038/s41567-018-0407-2}{2019 {\em Nat. Phys.\/} {\bf 15} 262--268 }

\bibitem{2009_PRL_Wang}
Wang F, Pollmann F and Vishwanath A \href{https://link.aps.org/doi/10.1103/PhysRevLett.102.017203}{2009 {\em Phys. Rev. Lett.\/} {\bf 102}(1) 017203} 

\bibitem{2009_PRB_Jiang}
Jiang H~C, Weng M~Q, Weng Z~Y, Sheng D~N and Balents L \href{https://link.aps.org/doi/10.1103/PhysRevB.79.020409}{2009 {\em Phys. Rev. B\/} {\bf 79}(2) 020409} 

\bibitem{2010_PRL_Heidarian}
Heidarian D and Paramekanti A \href{https://link.aps.org/doi/10.1103/PhysRevLett.104.015301}{2010 {\em Phys. Rev. Lett.\/} {\bf 104}(1) 015301} 

\bibitem{2021_PT_Akai}
Murtazaev A, Badiev M, Ramazanov M and Magomedov M \href{https://doi.org/10.1080/01411594.2021.1938047}{2021 {\em Phase Trans.\/} {\bf 94} 394--403 }

\bibitem{2017_PRB_Hiroyuki}
Chanlert P, Kurita N, Tanaka H, Kimata M and Nojiri H \href{https://link.aps.org/doi/10.1103/PhysRevB.96.064419}{2017 {\em Phys. Rev. B\/} {\bf 96}(6) 064419} 

\bibitem{2009_PRL_Chris}
Stock C, Chapon L~C, Adamopoulos O, Lappas A, Giot M, Taylor J~W, Green M~A, Brown C~M and Radaelli P~G \href{https://link.aps.org/doi/10.1103/PhysRevLett.103.077202}{2009 {\em Phys. Rev. Lett.\/} {\bf 103}(7) 077202} 

\bibitem{1995_PU_Dotsenko}
Dotsenko V~S \href{https://dx.doi.org/10.1070/PU1995v038n05ABEH000084}{1995 {\em Phys. Usp.\/} {\bf 38} 457} 

\bibitem{1994_HI_Bruce}
Gaulin B~D \href{1994 https://doi.org/10.1007/BF02069416}{{\em Hyperfine Interact.\/} {\bf 85} 159--171} 

\bibitem{1970_PRL_Leggett}
Leggett A~J \href{https://link.aps.org/doi/10.1103/PhysRevLett.25.1543}{1970 {\em Phys. Rev. Lett.\/} {\bf 25}(22) 1543--1546} 

\bibitem{2024_nature_sunpeijie}
Xiang J, Zhang C, Gao Y, Schmidt W, Schmalzl K, Wang C~W, Li B, Xi N, Liu X~Y, Jin H, Li G, Shen J, Chen Z, Qi Y, Wan Y, Jin W, Li W, Sun P and Su G \href{https://doi.org/10.1038/s41586-023-06885-w}{2024 {\em Nature\/} {\bf 625} 270--275} 

\bibitem{2025_PRB_Xu}
Xu X, Wu Z, Chen Y, Huang Q, Hu Z, Shi X, Du K, Li S, Bian R, Yu R, Cui Y, Zhou H and Yu W \href{https://link.aps.org/doi/10.1103/gsk8-1k9q}{2025 {\em Phys. Rev. B\/} {\bf 112}(12) 125163} 

\bibitem{2022_PNAS_JMSheng}
Sheng J, Wang L, Candini A, Jiang W, Huang L, Xi B, Zhao J, Ge H, Zhao N, Fu Y, Ren J, Yang J, Miao P, Tong X, Yu D, Wang S, Liu Q, Kofu M, Mole R, Biasiol G, Yu D, Zaliznyak I~A, Mei J~W and Wu L \href{https://www.pnas.org/doi/abs/10.1073/pnas.2211193119}{2022 {\em Proc. Natl. Acad. Sci.\/} {\bf 119} e2211193119} 

\bibitem{2024_PRL_zhu}
Zhu M, Romerio V, Steiger N, Nabi S~D, Murai N, Ohira-Kawamura S, Povarov K~Y, Skourski Y, Sibille R, Keller L, Yan Z, Gvasaliya S and Zheludev A \href{https://link.aps.org/doi/10.1103/PhysRevLett.133.186704}{2024 {\em Phys. Rev. Lett.\/} {\bf 133}(18) 186704} 

\bibitem{2024_arxiv_Chen}
Chen T, Ghasemi A, Zhang J, Shi L, Tagay Z, Chen Y, Chen L, Choi E~S, Jaime M, Lee M, Hao Y, Cao H, Winn B, Podlesnyak A~A, Pajerowski D~M, Zhong R, Xu X, Armitage N~P, Cava R and Broholm C \href{https://arxiv.org/abs/2402.15869}{(2024), arXiv:2402.15869 [cond-mat.str-el]}

\bibitem{2025_arxiv_cui}
Cui Y, Wu Z, Sun Z, Du K, Luo J, Li S, Yang J, Wang J, Zhou R, Chen Q, Kohama Y, Miyata A, Yang Z, Yu R and Yu W \href{https://arxiv.org/abs/2509.26151}{(2025), arXiv:2509.26151 [cond-mat.str-el]}

\bibitem{1998_PRL_Singh}
Singh R~R~P and Elstner N \href{https://link.aps.org/doi/10.1103/PhysRevLett.81.4732}{1998 {\em Phys. Rev. Lett.\/} {\bf 81}(21) 4732--4735} 

\bibitem{2018_PRB_Jozef}
Stre\ifmmode~\check{c}\else \v{c}\fi{}ka J, Kar\ifmmode~\check{l}\else \v{l}\fi{}ov\'a K, Baliha V and Derzhko O \href{https://link.aps.org/doi/10.1103/PhysRevB.98.174426}{2018 {\em Phys. Rev. B\/} {\bf 98}(17) 174426} 

\bibitem{2024_PRL_chengang}
Chen G~V \href{https://link.aps.org/doi/10.1103/PhysRevLett.133.136703}{2024 {\em Phys. Rev. Lett.\/} {\bf 133}(13) 136703} 

\bibitem{2020_PRB_zhongruidan}
Zhong R, Guo S, Nguyen L~T and Cava R~J \href{https://link.aps.org/doi/10.1103/PhysRevB.102.224430}{2020 {\em Phys. Rev. B\/} {\bf 102}(22) 224430} 

\bibitem{2024_CM_XuXianghan}
Xu X, Chen T, Wang H, Xie W, Broholm C~L and Cava R~J \href{https://pubs.acs.org/doi/full/10.1021/acs.chemmater.3c02980}{2024 {\em Chem. Mater.\/} {\bf 36} 7} 

\bibitem{2024_PRB_MeiJW}
Yue L, Lu Z, Yan K, Wang L, Guo S, Guo R, Chen P, Chen X and Mei J~W \href{https://link.aps.org/doi/10.1103/PhysRevB.109.214430}{2024 {\em Phys. Rev. B\/} {\bf 109}(21) 214430} 

\bibitem{2024_PRB_WangJ.F}
Li Z~R, Ouyang Z~W, Cao J~J, Wang L, Wang Z~X, Xia Z~C and Wang J~F \href{https://link.aps.org/doi/10.1103/PhysRevB.109.224413}{2024 {\em Phys. Rev. B\/} {\bf 109}(22) 224413} 

\bibitem{1992_jpsj_Kawamura}
Kawamura H \href{https://doi.org/10.1143/JPSJ.61.1299}{1992 {\em J. Phys. Soc. Jpn.\/} {\bf 61} 1299--1325} 

\bibitem{1982_RMP_FYWu}
Wu F~Y \href{https://link.aps.org/doi/10.1103/RevModPhys.54.235}{1982 {\em Rev. Mod. Phys.\/} {\bf 54}(1) 235--268} 

\bibitem{1950_ZETF_pomeranchuk}
Pomeranchuk I 1950 {\em Zh. Eksp. Teor. Fiz\/} {\bf 20} 16

\bibitem{1997_RMP_Lee}
Lee D~M \href{https://link.aps.org/doi/10.1103/RevModPhys.69.645}{1997 {\em Rev. Mod. Phys.\/} {\bf 69}(3) 645--666} 

\bibitem{1997_RMP_Richardson}
Richardson R~C \href{https://link.aps.org/doi/10.1103/RevModPhys.69.683}{1997 {\em Rev. Mod. Phys.\/} {\bf 69}(3) 683--690} 

\bibitem{2025_PRB_Xtao}
Chen J~L, Fan Z, Zhan B, Hu J, Liu T, Ji J, Wang K, Liao H~J and Xiang T \href{https://link.aps.org/doi/10.1103/1gx9-wcf6}{2025 {\em Phys. Rev. B\/} {\bf 112}(12) 125130} 

\bibitem{SM}
See Supplemental Material for the crystal structure and magnetic interactions, the low-temperature NMR spectra in the ordered phase, the NMR spectral simulation, the staggered magnetic moments in the UUD phase, and the theoretical modeling and the transition line between the UUD and the supersolid V phase, which include Refs.~\cite{2024_PRB_MeiJW,2024_PRB_WangJ.F,2025_PRB_Li}

\bibitem{2025_PRB_Li}
Li Z~R, Wang L, Cao J~J, Dong C, Tian Z~M, Wang Z~X, Xia Z~C, Nojiri H and Ouyang Z~W \href{https://link.aps.org/doi/10.1103/6csj-22vw}{2025 {\em Phys. Rev. B\/} {\bf 112}(9) 094456} 


\end{thebibliography}

\end{document}